\documentclass[aip,reprint]{revtex4-1}

\draft 

\usepackage[utf8]{inputenc}					
\usepackage{graphicx}					
\usepackage[ngerman,english]{babel}			
\usepackage{mathtools}					

\usepackage{wrapfig}
\usepackage{array}
\usepackage[table]{xcolor}
\usepackage{color}
\usepackage{subfigure}					
\usepackage[version=3]{mhchem}				

\usepackage{gensymb}					
\usepackage{float}  						

\begin{document}

\title{Measuring contact angle and meniscus shape with a reflected laser beam}%

\author{T. F. Eibach}
\affiliation{Max Planck Institute for Polymer Research, Ackermannweg 10, 55128 Mainz, Germany}
\author{D. Fell}
\affiliation{Max Planck Institute for Polymer Research, Ackermannweg 10, 55128 Mainz, Germany}
\affiliation{Center of Smart Interfaces, Technical University Darmstadt, 64287 Darmstadt, Germany}
\author{H. Nguyen}
\affiliation{Max Planck Institute for Polymer Research, Ackermannweg 10, 55128 Mainz, Germany}
\author{H. J. Butt}
\affiliation{Max Planck Institute for Polymer Research, Ackermannweg 10, 55128 Mainz, Germany}
\author{G. K. Auernhammer}
\email{auhammer@mpip-mainz.mpg.de}
\affiliation{Max Planck Institute for Polymer Research, Ackermannweg 10, 55128 Mainz, Germany}

\date{\today}

\begin{abstract}
Side-view imaging of the contact angle between an extended planar solid surface and a liquid is problematic. Even when aligning the view perfectly parallel to the contact line, focusing one point of the contact line is not possible. We describe a new measurement technique for determining contact angles with the reflection of a widened laser sheet on a moving contact line. We verified this new technique measuring the contact angle on a cylinder, rotating partially immersed in a liquid. A laser sheet is inclined under an angle $\varphi$ to the unperturbed liquid surface and is reflected off the meniscus. Collected on a screen, the reflection image contains information to determine the contact angle. When dividing the laser sheet into an array of laser rays by placing a mesh into the beam path, the shape of the meniscus can be reconstructed from the reflection image. We verified the method by measuring the receding contact angle versus speed for aqueous cetyltrimethyl ammonium bromide solutions on a smooth hydrophobized as well as on a rough polystyrene surface.
\end{abstract}


\maketitle

\section{Introduction}
\label{section_Introduction}
Contact angels are measured at the three phase contact line, where the three phases gas, solid and liquid meet each other. Many methods for measuring contact angles have been developed. They are normally based on measuring interfacial tensions or side-view imaging \citep[][pp. 31 - 91]{Good1979}. One method based on measuring surface tensions follows the work by Ludwig Wilhelmy \citep[][]{Wilhelmy1863} and is nowadays implemented in various commercial tensiometers. A vertically oriented plate is partially immersed in a liquid. The force necessary to hold this plate is measured. This force is composed of the weight of the plate, the buoyancy of the immersed part of the plate and the capillary force acting on the plate $F_{cap}=\gamma \cdot L \cdot \cos ( \theta )$. For a liquid with a known surface tension $\gamma$ and a plate with a known length of the contact line $L$, the contact angle can be calculated from this force measurement.\par
Methods based on imaging the contact line from the side have been used since the very beginning of the study of wetting phenomena \citep[][]{Young1805}. These methods depend on the optical resolution of the components. From side-view images, the contact angle can be measured directly. Numerous versions of this method exist and some of them are commercially available.
In further techniques sessile drops are used for measuring contact angles \citep[][]{Good1979}. From sessile drops on surfaces, images of the shape are made, where the contact angle can be determined.\par
An alternative of measuring the angle between the liquid surface and the solid surface directly with optical imaging is the tilting plate method \citep[][]{Adam1925}. A plate is immersed in the liquid and tilted until the surface of the liquid becomes completely flat. This can be checked precisely by side-view imaging or light reflection. \par
Graf et al. \citep[][]{Graf1998} were measuring contact angles in static equilibrium. They illuminated the three phase contact line with a laser, so that they received a reflection from the substrate and from the liquid. From the position of the reflections on a screen they calculated the contact angle.\par
Rio et al. were measuring contact angles of liquids, flowing down an inclined plane and leaving a dry patch inside \citep[][]{Rio2004} and of drops flowing down an inclined plane. \citep[][]{Rio2005} In both cases the contact angle was determined by having refractions of an illuminated laser. \par
Despite this wealth of methods, there are still situations in which contact angle measurements are challenging. Focusing on a specified point of the contact line is not possible. \par

In this article we present a new method to determine the dynamic contact angle and the shape of the meniscus from the reflection of a laser sheet on the substrate and the fluid-air interface. We show that contact angles and the meniscus shape can be measured and reconstructed with direct access to the contact line region and with a high temporal resolution.
\section{Reflection of a laser beam on a meniscus}
\label{section_Measuring_contact_angle}
In a setup similar to the one used by Fell et al. \citep[][]{Fell2011, Fell2011_2}, a cylinder rotated in a liquid and caused a deformation of the liquid surface. With this setup the authors could determine the velocity dependent contact angle with side view imaging. Here, we used a laser sheet to illuminate a line across the contact line and analyzed the reflection of this laser sheet. The reflections, collected on a screen, displayed a characteristic shape, from which the contact angle and the shape of the meniscus could be calculated.

\subsection{Experimental setup}
\label{subsection_Experimental_setup}
\begin{figure}[t]
	\centering
	\includegraphics[width=8.5cm]{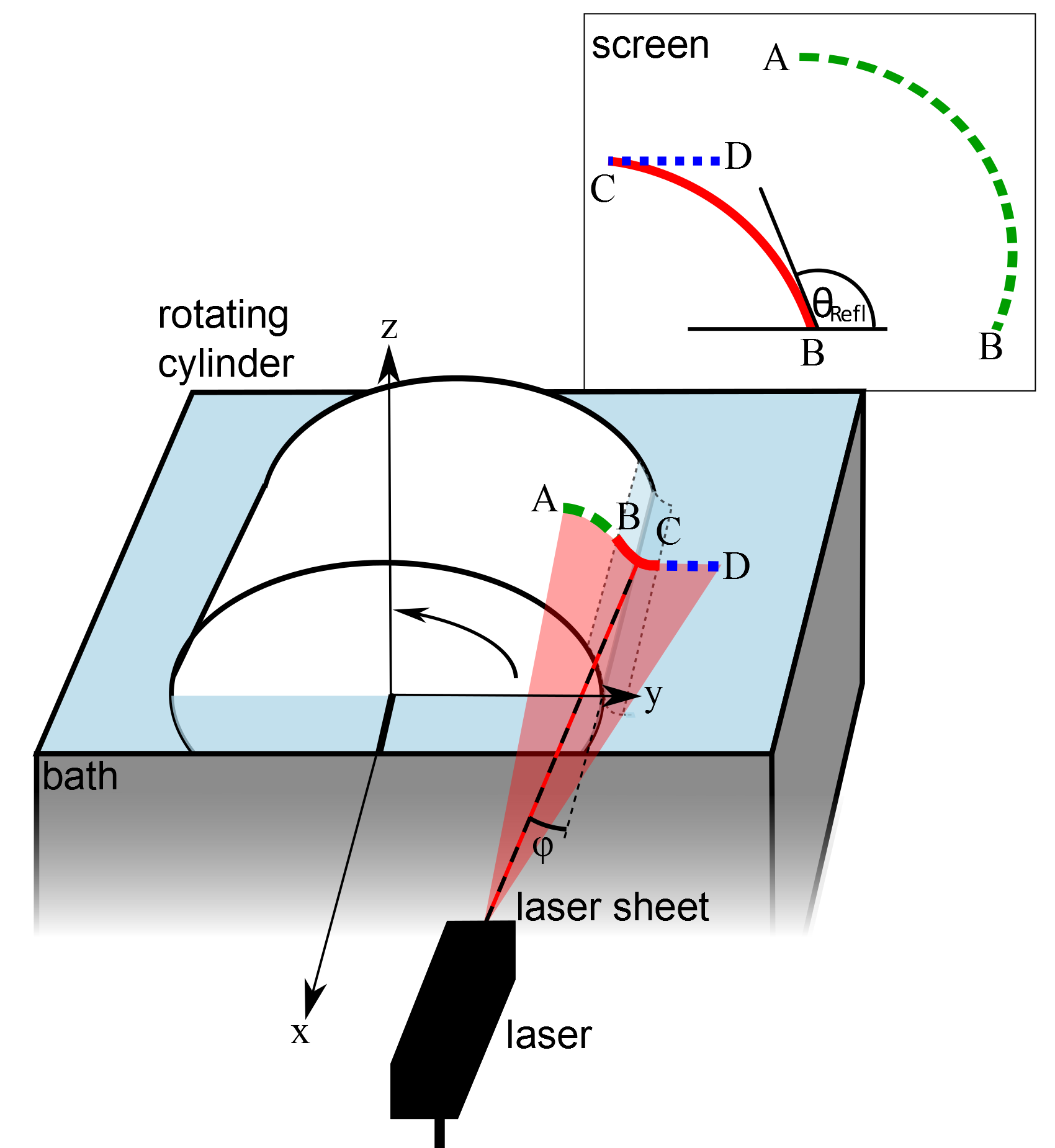}
	\caption[]{Schematic of the experimental setup. The cylinder was rotating in a bath filled with a liquid. A widened laser sheet with a wavelength of $\lambda = 635\,nm$ was inclined under an angle $\varphi$ to the liquid surface and was reflected on the cylinder (green dashed), the meniscus (red) and the unperturbed liquid surface (blue dotted). The reflection image was collected on a screen where the angle $\theta_{Refl}$ can be detected.}
	\label{Aufbau}
\end{figure}
\begin{figure}[t]
	\centering
	\includegraphics[width=5.5cm]{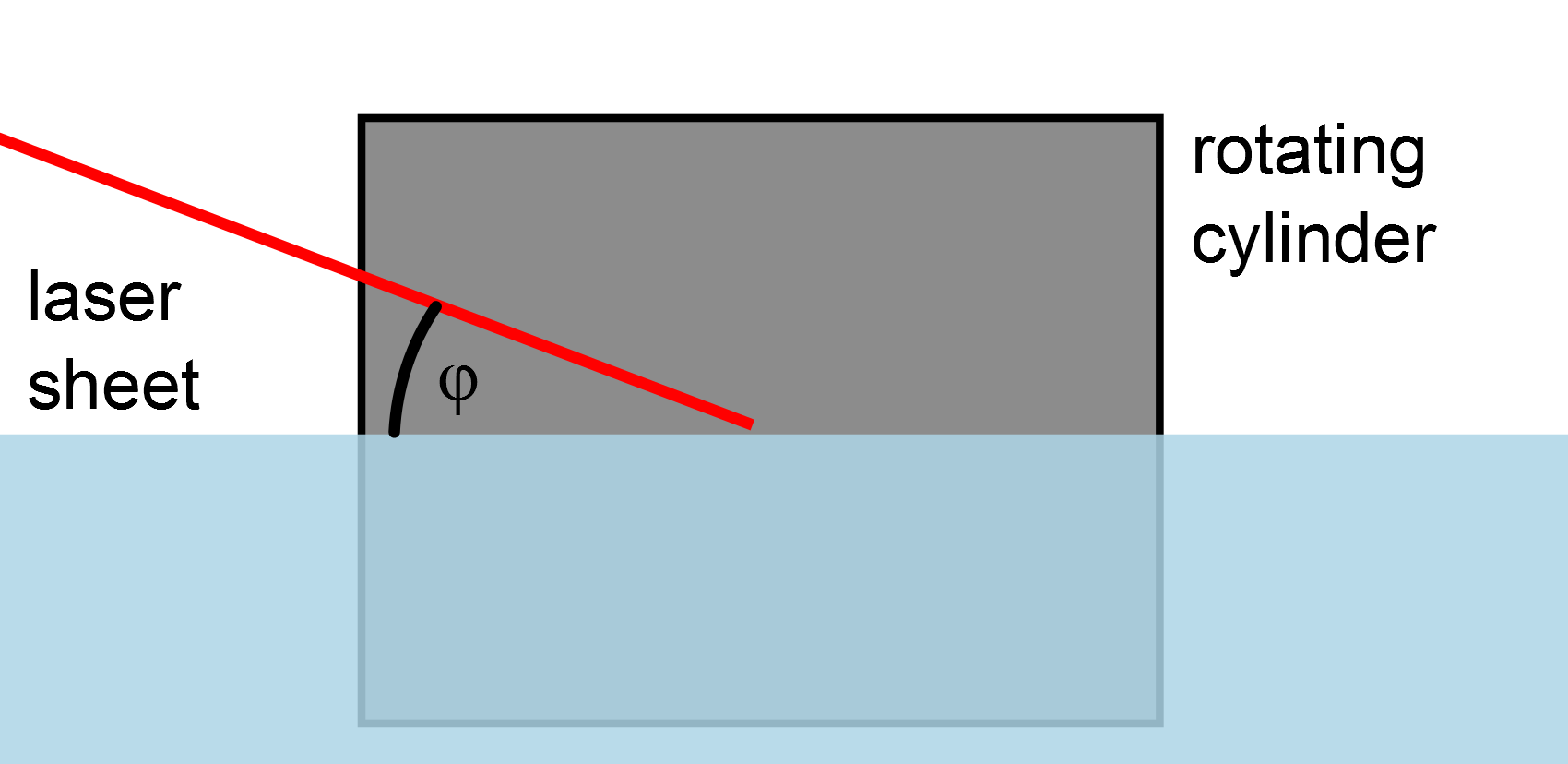}
	\caption[]{Schematic of the experimental setup. View from the side on the cylinder, rotating in the bath. The laser sheet is inclined under an angle $\varphi$ to the liquid surface.}
	\label{Aufbau_2}
\end{figure}\par
\begin{figure}[t]
	\centering
	\includegraphics[width=8.5cm]{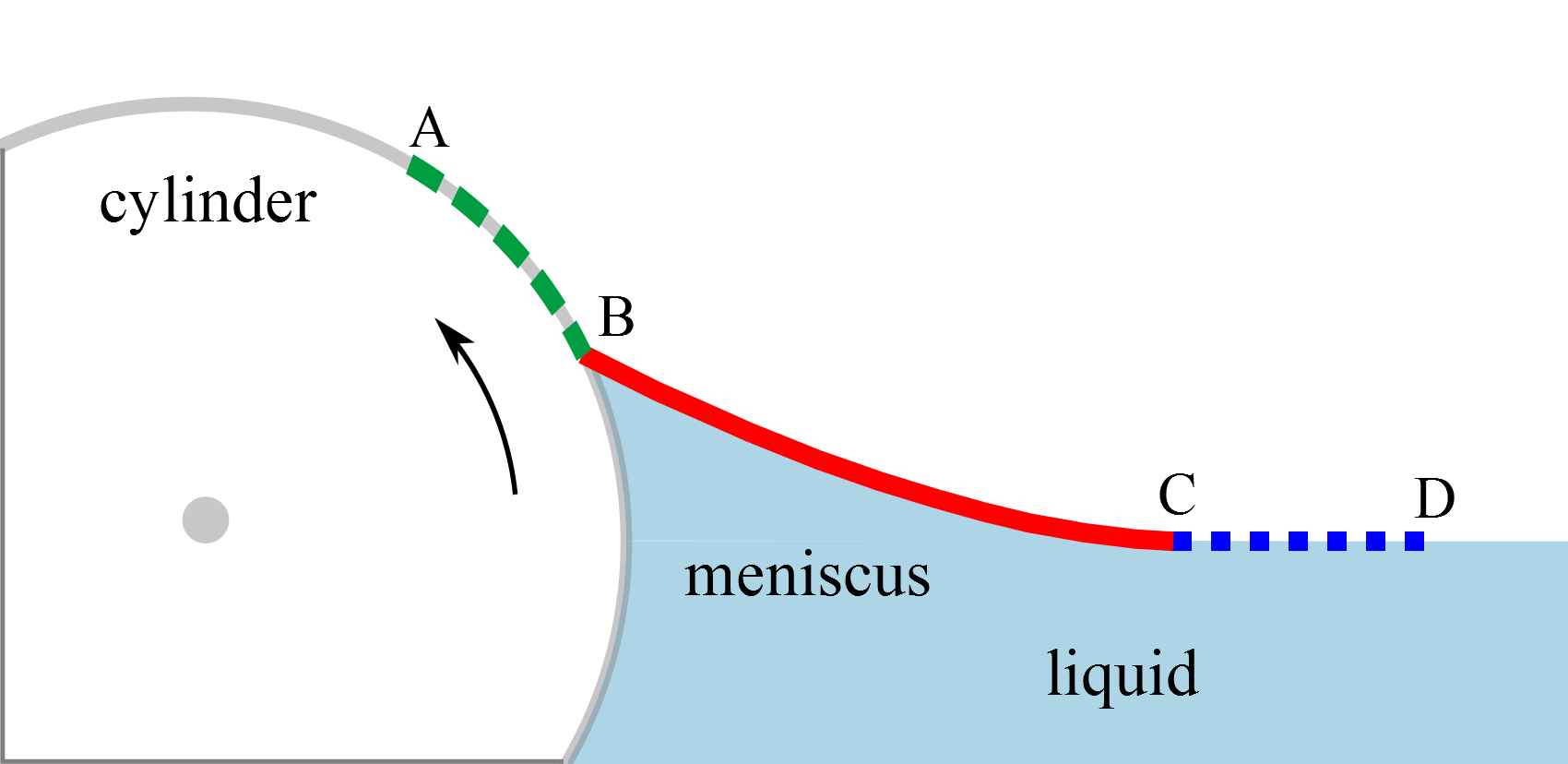}
	\caption[]{Schematic side-view of the rotating cylinder and the meniscus. The locations where the widened laser sheet is reflected are: the cylinder surface (from A to B), the meniscus (B to C), and the plane liquid surface (C to D).}
	\label{Reflektion_des_Lasers}
\end{figure}

The contact line between the cylinder and the liquid was illuminated by a laser (Lasiris SNF, Coherent, Santa Clara, CA 95006 USA) with a wavelength of $635\,nm$ and $10\,mW$ power. The laser beam was enlarged in one spatial direction by an optical lens generating a laser sheet with an opening angle of $\gamma = 10\,^{\circ}$ and a thickness of $d = 50 - 75\, \mu m$. The laser sheet was inclined under an angle $\varphi$ to the liquid surface where it hit the cylinder and the liquid surface (Fig. \ref{Aufbau_2}). The laser sheet was oriented such that the central ray of the laser sheet and the normal to the laser sheet span a plane that was parallel to the x-z-plane (Fig. \ref{Aufbau}). To describe the reflection of the laser sheet, it is convenient to distinguish three different regions for the reflection (Fig. \ref{Reflektion_des_Lasers}): the cylinder surface (from A to B), the meniscus (B to C), and the plane liquid surface (C to D). The reflections hit a screen that was positioned behind the bath. The reflections were recorded with a camera (Olympus, i-Speed LT equipped with a SIGMA AF 50/2.8 DG Macro objective) with a maximum frame rate of $2000\, fps$.\par
To reconstruct the meniscus profile, the laser sheet was split up in single laser beams by placing a steal mesh (Haver \& Boecker, 59302 Oelde, Germany) in the laser sheet. The mesh consisted of woven steel wires with a thickness of $63\, \mu m$ and an opening of $100\, \mu m$, so that it had a distance between the openings of the mesh of $163\, \mu m$. This led to a group of individual rays being reflected from the meniscus.\par

The setup was based on a bath, which was about $9\,cm$ in width, $15\,cm$ in depth, and $15\,cm$ in height, made out of polyvinyl chloride (PVC). In the bath, a horizontally oriented cylinder was rotating, driven by a motor with velocities ranging from $0.01\,cm/s$ up to $100\,cm/s$. The cylinder was made of stainless steel with a radius of $6\,cm$ and a length of $5\,cm$ (Fig. \ref{Aufbau}). Two different surfaces of the cylinder were used: the polished stainless steel surface and thin glass plates ($50\,mm \times 200\,mm \times 70\, \mu m$, kindly provided by Gerhard Menzel GmbH, 38116 Braunschweig, Germany) that were bent and glued on the cylinder. To hydrophobize the surfaces they were coated. The steel cylinder was coated with polystyrene. To do so, it rotated at a speed of $8\,cm/s$ in a $0.8\,wt\, \%$ solution of polystyrene (PS) ($M_w = 163\,kg/mol$) in tetrahydrofuran. Because of the evaporation of the solvent, a polystyrene film remained on the cylinder.\citep[][]{Fell2011} 
To coat the glass plates with hexamethyldisilazane (HMDS), a solution of $2.5\,wt\, \%$ of HMDS ($M_w = 0.161\,kg/mol$) in toluene was used. When coating with HMDS the evaporation played no role, because it was bond chemically to the glass. The coatings were dried for 16 (PS) and 3 hours (HMDS) in the oven at $60\,^{\circ}C$, respectively, before being used for the wetting experiment. The roughnesses were measured with an atomic force microscope (AFM). For silanized glass the roughness was $\approx 1\, nm$ (RMS) on a surface of $25\mu m \times 25\mu m$. For PS on steel it was $167 \pm 38\,nm$ (RMS) on a surface of $50\mu m \times 50\mu m$.\citep[][]{Fell2012}
The bath was filled with liquid, so that the cylinder was more than half immersed in it. 
Two liquids were used in the presented experiments: Milli-Q water and aqueous solutions of the cationic surfactant CTAB (cetyltrimethylammonium bromide; $\ce{CH3(CH2)15N+(CH3)3BR-}$, critical micelle concentration (cmc) $= 1\,mmol/l$). All experiments were carried out in a water saturated atmosphere. For verification, contact angles were recorded with the side view method described by Fell et al. \citep[][]{Fell2011}.

\subsection{Modeling with Geometrical Optics}
\label{subsection_theoretical}
We used geometrical optics to follow the beam path of the individual rays of the laser sheet. Deduced formulas relate the shape of the reflection to the contact angle and the shape of the meniscus. The used coordinate system is given in Fig. \ref{Aufbau}. Additional parameters are: the position of the laser $\vec{p} = \left( p_1, p_2, p_3 \right)$, the angle $\varphi$, under which the laser sheet is inclined to the liquid surface, the opening angle $\gamma$ of the laser sheet, the radius $r$ and the position of the cylinder, the reference point $\vec{q} = \left( q_1, q_2, q_3 \right)$ of the screen and its normal vector $\vec{n_S}$. Also the position of the unperturbed liquid surface must be known. When knowing these parameters, the laser sheet is represented by an array of rays (straight lines)
\begin{equation}
g_i: \vec{x} = \vec{p} + \lambda \vec{a_i}
\label{Gleichung_1}
\end{equation}
with the scaling parameter $\lambda$ and the direction vector
\begin{align}
\vec{a_i} = \left(\begin{array}{c} a_{i1} \\ a_{i2} \\ a_{i3} \end{array}\right)
 &= \left(\begin{array}{c} -\cos( \varphi ) \cdot \cos( \epsilon_i ) \\ \sin( \epsilon_i ) \\ -\sin( \varphi ) \cdot \cos( \epsilon_i ) \end{array}\right) \notag \\ &\approx \left(\begin{array}{c} -\cos( \varphi ) \\ \epsilon_i \\ -\sin( \varphi ) \end{array}\right)
\label{Gleichung_13}
\end{align}
that depends on the array index $i$. Here the last approximation is only valid for small opening angles of the laser sheet. $\epsilon$ is the angle between a single ray and the x-axis. In further calculations we focus on one ray and omit the index $i$ for clarity. Details of the calculation can be found in the Appendix \ref{section_Calculation2}.

\subsection{Calculation of the contact angle}
\label{subsection_calculation}
\begin{figure}[t]
  \centering
  \subfigure[]{
    \label{Beispiel_1_1}
    \includegraphics[width=4cm]{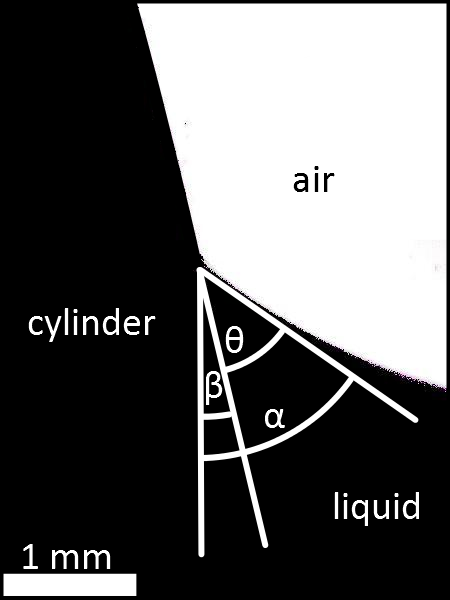}
  }
  \subfigure[]{
    \label{Beispiel_1_2}
    \includegraphics[width=4cm]{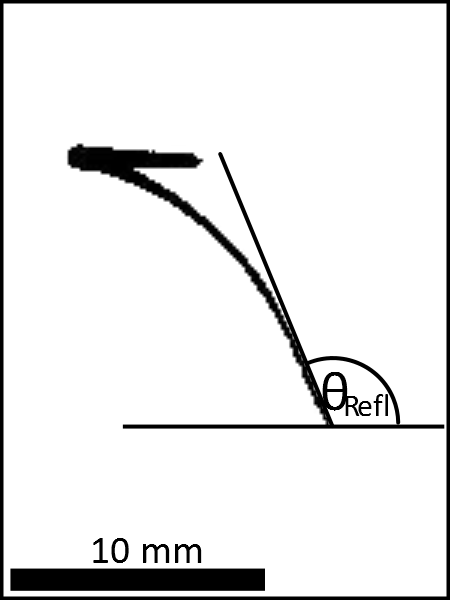}
  }
  \caption[]{(a) Side-view on the cylinder and the meniscus. With the reflection method the inclination angle $\alpha$ of the meniscus is calculated. Subtracting the inclination angle $\beta$ of the surface at the contact line, one receive the contact angle $\theta$. (b) From the reflection image the angle $\theta_{Refl}$ was determined for calculating the contact angle $\theta$.}
  \label{Beispiel_1}
\end{figure}

The slope $m_{Refl}$ of the meniscus reflection is related to the slope of the meniscus $m$ at the reflection point (Appendix \ref{section_Calculation2})
\begin{equation}
m_{Refl} = \frac{2m}{1-m^2}
\label{Gleichung_8}
\end{equation}
At the end point of the reflection, e.g., at the three phase contact line, we can rewrite this relation with the identities $m=-\frac{1}{\tan (\alpha )}$ and $m_{Refl}=\tan (\theta_{Refl} )$ (Fig. \ref{Beispiel_1}): 
\begin{equation}
\alpha = \frac{1}{2} \theta_{Refl}.
\label{Gleichung_9}
\end{equation}
For getting the contact angle $\theta$, the angle $\beta$ between the vertical line and the cylinder surface at the contact line must be taken into account
\begin{equation}
\theta = \alpha - \beta = \frac{1}{2} \theta_{Refl} - \beta.
\label{Gleichung_10}
\end{equation}
We analyzed the reflection images with a self written program in MATLAB. For an accurate positioning of the reflection, the position of all bright pixels belonging to the reflection were averaged horizontally. The resulting average shape of the reflection was fitted with a smooth function. At the end point of the reflection, the angle between the tangential and the horizontal line defines the angle $\theta_{Refl}$. Using Eq. (\ref{Gleichung_10}), we deduced the contact angle $\theta$ (Fig. \ref{Beispiel_1}). $\beta$ was easily measured directly from a side-view image of the cylinder or calculated from the filling height. For automatical analysis, a video with around $100$ frames was taken and the contact angles were calculated for every image. In the experimental part below, the contact angles for one video are averaged. The error represents the standard deviation.

\subsection{Reconstruction of the meniscus profile}
\label{subsection_reconstruction}
\begin{figure}[t]
  \centering
  \subfigure[]{
    \label{Meniscus_reconstruction_2}
    \includegraphics[width=4cm]{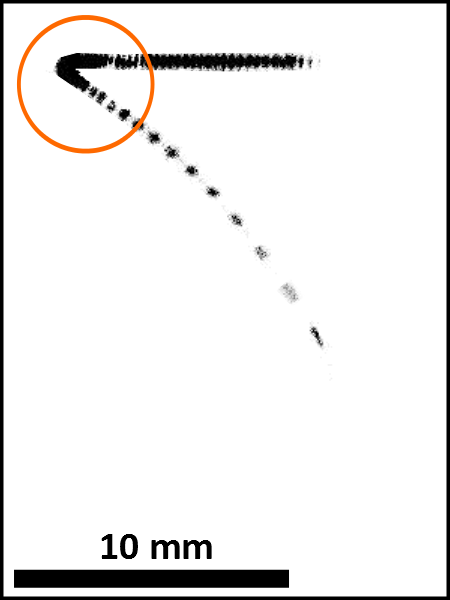}
  }
  \subfigure[]{
    \label{Meniscus_reconstruction_1}
    \includegraphics[width=4cm]{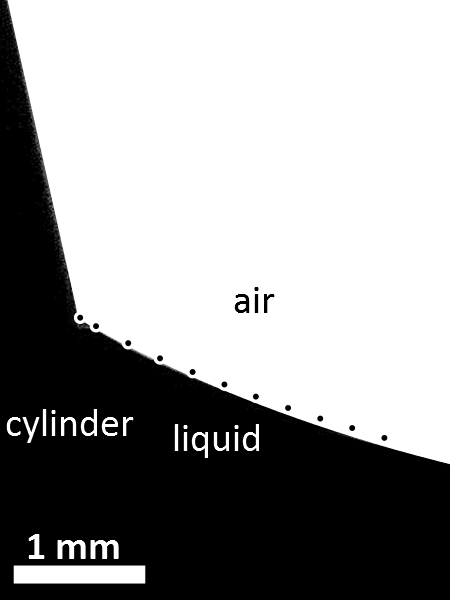}
  }
  \caption[]{(a) Dotted reflection image with every dot belonging to a ray originating from one opening in the mesh. Two neighboring rays have an angular distance of $e \approx \arctan \left(\frac{b}{l_1} \right) \approx 0.0048\,^{\circ}$, with the distance between the openings of the mesh $b$ and the distance $l_1$ between laser and mesh. The rays reflected in the vicinity of the contact line were easily identifiable. Far away from the contact line the dots could not be separated and were not included in the reconstruction of the meniscus shape. (b) Side-view image superimposed with the reconstructed shape of the meniscus (black dots with white surrounding).}
  \label{Meniscus_reconstruction}
\end{figure}

For reconstructing the meniscus shape from the reflection, the laser sheet was split into single rays. They were reflected and generated a dotted reflection image. Every dot originated from the ray from one opening in the mesh (Fig. \ref{Meniscus_reconstruction_2}). The theoretical analysis did not change in this case. The angular distance between the rays was given by the distance between the openings in the mesh and the position of the mesh in the setup.
\par 
For calculating the contact angles, the reflections of the single rays on the screen must be located. This was done with the particle tracking-algorithm by Crocker, Grier, and Weeks \citep[][]{Crocker1996}, transferred to MATLAB by D. Blair and E. Dufresne \citep[][]{MATLAB2012}. The dots reflected close to the point C (and further away from the contact line) could not be separated in the image analysis and were not included in the reconstruction of the meniscus shape (orange circle in Fig. \ref{Meniscus_reconstruction_2}). The slope between two dots in the reflection image was determined to $m_{Refl}=\frac{\Delta z}{\Delta y}$. Eq. (\ref{Gleichung_8}) yields to the slope $m$ of the meniscus segment between the reflection points of these two rays. The distance on the meniscus between the central and another ray was calculated from the distance between laser and mesh $l_1$, between mesh and meniscus $l_2$ and the distance between the openings of the mesh $b$ to
\begin{equation}
c= \tan \left( \arctan \left( \frac{b \cdot (j-0.5)}{l_1} \right) \right) \cdot \left( l_1 + l_2 \right).
\label{Gleichung_14}
\end{equation}
Here, $j$ describes the number of rays away from the central ray. To reconstruct the meniscus shape, we first determined the slope of all its segments. The shape of the meniscus was calculated as the piecewise linear function of these segments. Starting from the known y- and z-coordinates of the contact line, the shape was composed by calculating the length and the slope of each segment. The length of the first segment was chosen to be half of the following segments. For the latter ones the length of the segments is given by Eq. (\ref{Gleichung_14}). The end points of the segments are compared to the meniscus shape in Fig. \ref{Meniscus_reconstruction_1}.

\section{Testing the method}
\label{section_verification}
We compared contact angles measured with our reflection method with contact angle measurement conventionally in side view
\citep[][]{Fell2011}. Since the optical paths of both setups overlap, the experiments could not be performed simultaneously. The experimental data was taken in the same conditions but in two independent experiments. The contact angle was varied either by changing the dewetting speed or by adding surfactant to the wetting liquid (water in our case).

\subsection{Contact angle measurement}
\label{subsection_Contact_angle}

\begin{figure}[t]
	\centering
	\includegraphics[width=8.5cm]{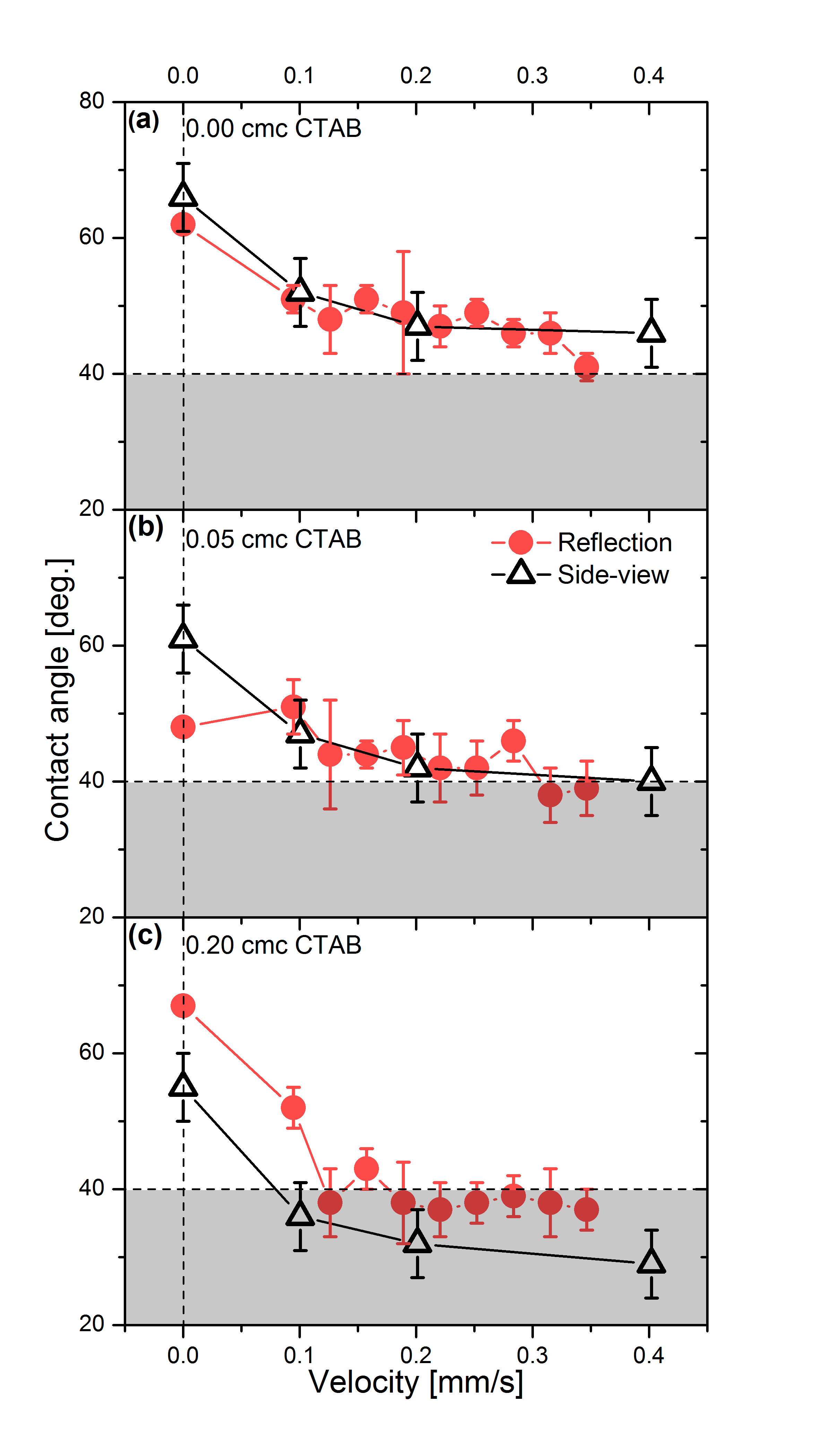}
	\caption[]{Contact angles vs. dewetting velocity,  measured with the reflection method (red circles) and the side-view method (black triangles) from the work by Fell \citep[][]{Fell2012}. Dewetting of a glass surface hydrophobized with HMDS. Contact angles were measured in solutions of Milli-Q water and the surfactant CTAB with concentrations (a) $0.00$ cmc CTAB, (b) $0.05$ cmc CTAB, (c) $0.20$ cmc CTAB. From the filling height of the water bath we calculated and confirmed with side-view imaging $\beta = (5 \pm 1)\,^{\circ}$. The horizontal dotted line at $40\,^{\circ}$ and the grey-shaded area indicate the minimal limitation for measurable contact angels with the reflection method. For details see section \ref{subsection_limitation_CA}.}
	\label{Glas_CTAB_Kombination}
\end{figure}
Fig. \ref{Glas_CTAB_Kombination}(a) shows the receding contact angle for pure water on smooth, hydrophobized glass as a function of the velocity. The contact angles decreased when the rotating velocity of the cylinder increased. For increasing CTAB concentration $\theta$-vs-$v$ decreased more steeply compared to pure water (Fig. \ref{Glas_CTAB_Kombination}(b), (c)). Some data points deviated from this general tendency. We attribute this to imperfections in the coating of the glass plate. The error bar (the standard deviation of the measured distribution of contact angles) was typically $\pm 5\,^{\circ}$.\par
Comparing the results obtained with the reflection method with previous results obtained in side view \citep[][]{Fell2012} showed good agreement. Differences were only $\pm 3\,^{\circ}$, i.e., with the error bar of the measurement. However, at a concentration of $0.20$ cmc CTAB the contact angle for the reflection method reached a plateau at a velocity of $0.13\,mm/s$. For higher velocities the measured contact angle remained nearly constant with fluctuations of the data points of $\pm 6\,^{\circ}$. The constancy arises from the lower limitation of contact angels, measured with the reflection method. The contact angles for the side-view method were decreasing over the whole range of velocities (see section \ref{subsection_limitation_CA} for the explanation).\par

\begin{figure}[t]
	\centering
	\includegraphics[width=8.5cm]{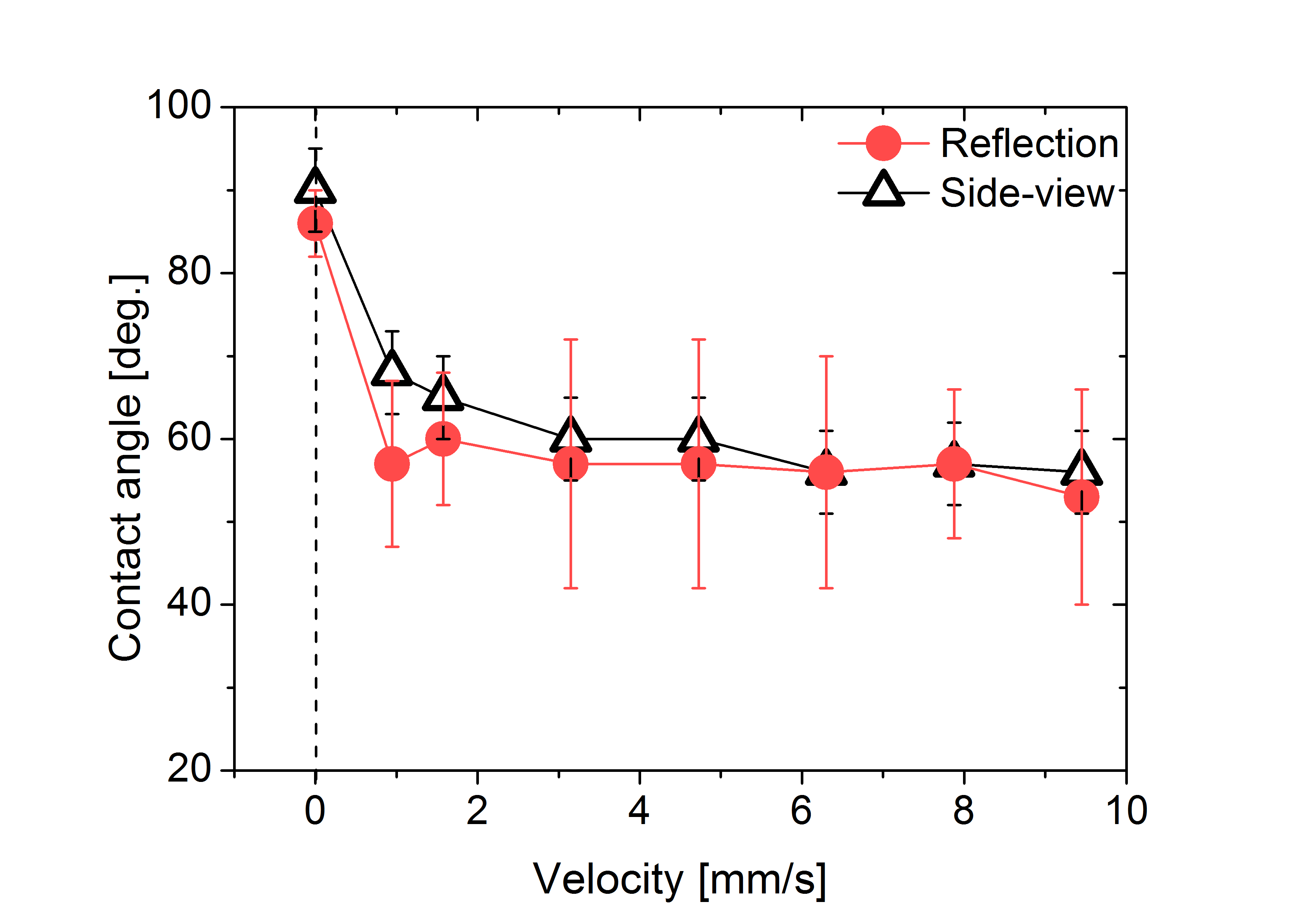}
	\caption[]{Contact angles vs. dewetting velocity, measured with the reflection method (red circles) and the side-view method (black triangles) for pure water on the steel cylinder coated with polystyrene.}
	\label{Steel_water_Reflexion}
\end{figure}
For steel cylinders coated with polystyrene we also found good agreement between contact angles measured with the reflection mode and one measured in side view by Fell et al. \citep[][]{Fell2011} (Fig. \ref{Steel_water_Reflexion}) .\par
In most cases the difference between both methods was less than $5\,^{\circ}$. However, in contrast to the smooth surface, the error bar of the reflection technique on the rough polystyrene was significantly larger than for the side-view technique (see section \ref{subsection_roughness} for a detailed discussion).

\subsection{Shape of the meniscus}
\label{subsection_reconstruction2}
A mesh with a distance between the openings of $b=163\, \mu m$ was positioned in the laser sheet in a distance of $l_1=19.4\,cm$ to the laser and a distance of $l_2=8.8\,cm$ to the meniscus. We positioned the laser that the central ray of the laser array was reflected on the contact line.\par
The slopes $m_{Refl}$ and $m$ are slopes of segments. Since we make a piecewise linear reconstruction of the meniscus shape, errors in determining the slope accumulate with increasing distance from the contact line (Fig. \ref{Meniscus_reconstruction_1}). 
Still, the real meniscus profile and the calculated one are in good agreement. Since the laser sheet was inclined under an angle $\varphi < 90\,^{\circ}$ with respect to the free water surface, the x-coordinates of the reflection points vary. In case that the reflections are done in the center of the cylinder, the profile can be supposed being independent of the x-coordinate.\par
The resolution of the reconstructed meniscus shape is given by the distance between two reflection points on the meniscus and can be raised when using a mesh with a smaller gap distance. In order to keep a good resolution of the individual spots on the screen the distance to the screen should be increased simultaneously.

\section{Discussion}
\label{section_results}

\subsection{Range of measurable contact angles}
\label{subsection_limitation_CA}
\begin{figure}[t]
	\centering
	\includegraphics[width=6.5cm]{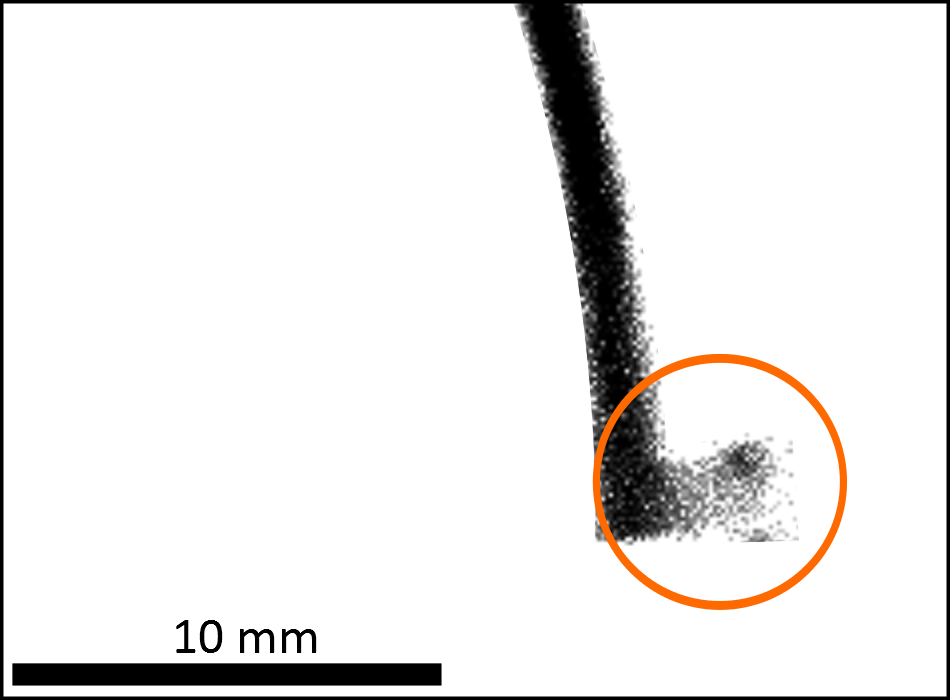}
	\caption[]{For an inclination angle $\alpha < 45\,^{\circ}$ the reflection from the meniscus is reflected from the water surface (orange circle).}
	\label{Reflection_water_surface}
\end{figure}

With the reflection method developed in this work only a limited range of contact angles can be measured. The lower limit is at an inclination angle of the liquid surface of $\alpha \approx 45\,^{\circ}$ (Eq. (\ref{Gleichung_9})). At inclination angles $\alpha < 45\,^{\circ}$, the laser beam is reflected onto the plane liquid surface. Here, the rays are reflected again from the liquid surface or enter the liquid (Fig. \ref{Reflection_water_surface}). Since the liquid interface is almost never completely flat, this reflection cannot be used for the analysis.\par
If the cylinder surface is not vertical at the contact line, $\alpha$ has to be corrected by the inclination angle $\beta$ of the cylinder surface to obtain the contact angle (Eq. (\ref{Gleichung_10})). In the experiments discussed above, the inclination angle of the cylinder surface was $\beta = (5 \pm 1)\,^{\circ}$. That led to the lower limit in this setup of $\theta \approx 40\,^{\circ}$. This limit is the reason for the plateau for contact angles at $0.20$ cmc CTAB (Fig. \ref{Glas_CTAB_Kombination} (c)).\par
The upper limit is at an inclination angle of $\alpha \approx 90\,^{\circ}$. In this case, the laser beam is reflected onto the cylinder surface. Correcting again for the inclination of the cylinder surface at the contact line led to an upper limit of $\theta \approx 85\,^{\circ}$ for the present setup.\par
To overcome these limitations, a change in geometry would be necessary. Using a laser sheet with a central no longer in the x-z-plane but rotated towards the y-axis should allow for a different range of inclination angles (also covering $0\,^{\circ} < \alpha < 45\,^{\circ}$. For contact angles beyond $90\,^{\circ}$, the laser sheet would have to pass through the liquid and be reflected from the liquid side at the liquid surface. A detailed analysis of these geometries is, however, beyond the scope of the present work.

\subsection{Relevant part of the meniscus for the determination of the contact angle}
\label{subsection_resolution}
\begin{figure}[t]
	\centering
	\includegraphics[width=8.5cm]{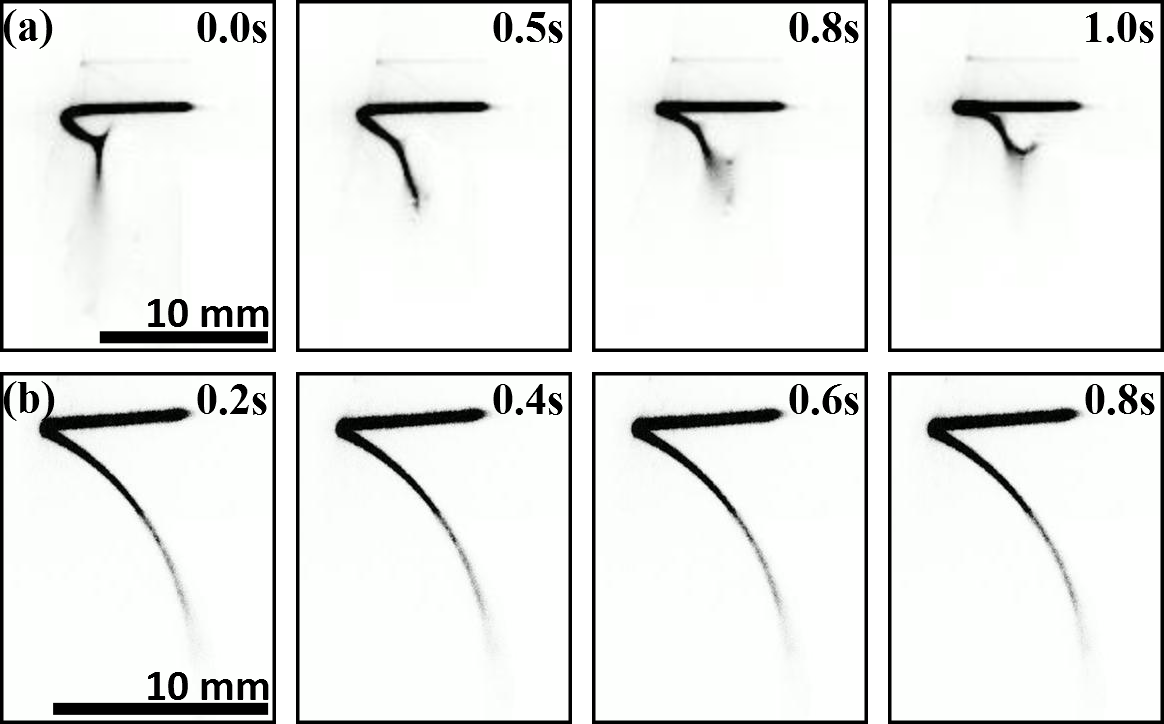}
	\caption[]{Movement of the meniscus reflection on the rough steel surface coated with PS (a) and the smooth glass surface coated with HMDS (b), rotating in pure water at a velocity of $0.35\,mm/s$. The meniscus of the rough steel surface is moving more than the meniscus on the smooth glass.}
	\label{Movement_Reflection}
\end{figure}
In the side-view method used by Fell et al. \citep[][]{Fell2011} the contact angle was calculated with an algorithm, in which the meniscus profile is extrapolated towards the contact line (Fig. \ref{Beispiel_1_1}). Typically the relevant data for the extrapolation is taken at a distance of about $200-400\,\mu m$ to the contact line. A lower extrapolation distance is not possible because of the optical resolution of the setup. Changes of the meniscus that are closer to the contact line, do not influence the measurement of the contact angle.\par
An advantage of the reflection method is that one analyzes the region close to the contact line, because that region is enlarged. From comparisons of the side-view imaging and the reconstruction of the shape of the meniscus, the main contribution comes from a region $30-60\,\mu m$ around the contact line. 
The resolution is also limited by the thickness of the laser sheet at the meniscus, the thicker the laser sheet, the lower the resolution.\par
Compared to the side-view technique, the reflection method is more sensitive for fluctuations and irregularities in the shape of the meniscus.

\subsection{Influence of surface roughness on the contact angle measurement}
\label{subsection_roughness}
For every velocity a video with around 100 frames is taken and the contact angle is determined for every frame. The mean of all contact angles in a video is the statistic contact angle and the standard deviation defines the error bar. Therefore the error bar is a measure of the fluctuations of the contact angle. Comparing the results for both surfaces discussed in section \ref{section_verification}, a difference in the error bars was obvious. On the steel surface with a roughness of $167 \pm 38\,nm$ the average of the error bars was $\pm 11\,^{\circ}$, being caused by big fluctuations of the contact angles (Fig. \ref{Movement_Reflection}(a)). On the glass surface with a roughness of $\approx 1\, nm$ there were less fluctuations (Fig. \ref{Movement_Reflection}(b)) with an average of the error bars of $\pm 3.5\,^{\circ}$.\par
The rougher the surface, the higher the error of the contact angle. This means, that the error bar is at least a qualitative measure for the surface roughness.

\section{Conclusion and Outlook}
Contact angles can be measured at a rate up to 2000 fps in a range of $45\,^{\circ}$ to $90\,^{\circ}$ with the new reflection method. Advantages of the reflection method are that contact angle of planar surfaces or cylinders can be measured close to the contact line. Optical elements such as lenses are not required. Therefore the working distance can be larger, allowing the application in large setups. The error bar of the contact angles include information of the roughness of a surface analyzed.\par
It is possible to measure contact angles in opaque fluids, since the reflections take place outside of the fluid phase. It is possible to reconstruct the profile of the whole meniscus from the reflection. This can be done by splitting the laser sheet into an array of separated laser rays. Such an array of laser rays can be generated by placing a mesh into the beam path of the laser sheet. The reconstruction permits to analyze the change of the meniscus with a high time resolution for different surface roughnesses and velocities. With a laser coming from the gas phase a contact angle between $45\,^{\circ}$ and $90\,^{\circ}$ can be measured. To cover the full range of contact angles between $0\,^{\circ}$ and $90\,^{\circ}$, slight modifications in the set-up would be necessary.
\appendix
\section{Calculation of the contact angle}
\label{section_Calculation2}
We used geometrical optics to follow the beam path and finally deduced a formula that relates the data from the reflection to the contact angle and the shape of the meniscus. The coordinate system is given in Fig. \ref{Aufbau}. The input parameters are: the position of the laser $\vec{p} = \left( p_1, p_2, p_3 \right)$, the angle $\varphi$, under which the laser sheet is inclined to the liquid surface, the opening angle $\gamma$ of the laser sheet, the radius $r$ and the position of the cylinder, the reference point $\vec{q} = \left( q_1, q_2, q_3 \right)$ of the screen and its normal vector $\vec{n_S}$. Also the position of the unperturbed liquid surface must be known. The laser sheet is represented by an array of straight lines
\begin{equation}
g_i: \vec{x} = \vec{p} + \lambda \vec{a_i}
\label{Gleichung_1}
\end{equation}
with the parameter $\lambda$ and the direction vector
\begin{align}
\vec{a_i} = \left(\begin{array}{c} a_{i1} \\ a_{i2} \\ a_{i3} \end{array}\right)
 &= \left(\begin{array}{c} -\cos( \varphi ) \cdot \cos( \epsilon_i ) \\ \sin( \epsilon_i ) \\ -\sin( \varphi ) \cdot \cos( \epsilon_i ) \end{array}\right) \notag \\ &\approx \left(\begin{array}{c} -\cos( \varphi ) \\ \epsilon_i \\ -\sin( \varphi ) \end{array}\right)
\label{Gleichung_2}
\end{align}
$\vec{a_i}$ depends on the array index $i$. Here $\epsilon$ is the angle between a single ray and the x-axis. In further calculations we focus on one ray and omit the index $i$ for better readability. Since $\epsilon < 5\,^{\circ}$ we used the small angle approximation $\sin( \epsilon ) \approx \epsilon$ and $\cos( \epsilon ) \approx 1$ in Eq. (\ref{Gleichung_2}).\par
To relate the slope of the reflected curve on the screen to the contact angle and the shape of the meniscus we used the following approach. For any given pair of two rays in the laser sheet there are tangential planes to the reflecting surface at the reflection point. Assuming an independency of the meniscus shape from the x-coordinate, these tangential planes can be described by equations of the type $z=m\cdot y+d$, with the local slope of the surface $m$ and an offset $d$. From the intersection with the corresponding ray $g$ we received the parameter
\begin{equation}
\lambda = \frac{d+mp_2-p_3}{-ma_2+a_3}.
\label{Gleichung_3}
\end{equation}
and the intersection point 
\begin{equation}
\vec{s_{1}}=\vec{p} + \frac{d+mp_2-p_3}{-ma_2+a_3} \cdot \vec{a}.
\label{Gleichung_11}
\end{equation}
The reflection of the ray at the tangential plane can be considered as a mirroring at the straight line $h:\vec{x} = \vec{s_1} + \mu \cdot \vec{n}$, with the normalized normal vector $\vec{n} = \left( 0, -m, 1 \right)$ of the reflective plane. The mirroring of any point $\vec{p}$ on $g$ is given as
\begin{equation}
\vec{p'} = \vec{p} + 2\cdot \left( \vec{s_p} - \vec{p} \right) =\vec{p} + 2  \lambda \cdot \left( \vec{a} - \frac{\vec{a} \cdot \vec{n}}{m^2+1} \cdot \vec{n} \right).
\label{Gleichung_4}
\end{equation}
$\vec{s_p}$ is the intersection point between the straight line $h$ and the plane through $\vec{p}$, parallel to the reflective plane. The reflected ray then is given by
\begin{equation}
g': \vec{x} = \vec{p'} + \nu \cdot \left(\vec{p'} - \vec{s_1} \right) 
\label{Gleichung_5}
\end{equation}
that was hitting the screen. It is described by the equation
\begin{equation}
E= \left( \vec{q} - \vec{x} \right) \cdot \vec{n_S},
\label{Gleichung_6}
\end{equation}
with $\vec{n_S} = \left( -1, 0, 0 \right)$. The intersection point between $g'$ and $E$ resultes in
\begin{equation}
\vec{s_2} = \left(\begin{array}{c} s_{21} \\ s_{22} \\ s_{23} \end{array}\right) = \vec{s_1} + \frac{q_1-s_{11}}{p_{1}'-s_{11}} \cdot \left(\vec{p'} - \vec{s_1} \right)
\label{Gleichung_7}
\end{equation}
seen on the screen.\par
For calculating the contact angle from the reflection image, two reflection points on the screen are considered that originate from two neighboring rays. The slope between both reflection points on the screen is $m_{Refl}= \frac{\Delta z}{\Delta y}$, with the difference of the $z$-coordinates $\Delta z$ and the difference of the $y$-coordinates $\Delta y$ on the screen. From Eq. (\ref{Gleichung_2}) we know that only $a_2$ is varying between the rays of a laser sheet; $a_1$ and $a_3$ remain constant. It is also assumed, that the slope $m$ and the offset $d$ of the meniscus remain constant. This leads to the slope
\begin{equation}
m_{Refl} = \frac{\Delta z }{\Delta y} = \frac{\Delta s_{23}(a_2)}{\Delta s_{22}(a_2)} = \cdot \cdot \cdot = \frac{2m}{-m^2+1}
\label{Gleichung_12}
\end{equation}
of the reflection on the meniscus in dependency to the slope $m$ of the meniscus.


\begin{acknowledgments}
We thank Stefan Geiter for the technical support. We acknowledge support by Gerhard Menzel GmbH (38116 Braunschweig, Germany) by providing the glass  plates. TFE acknowledges financial support by the Max Planck Society.
\end{acknowledgments}

\bibliography{Literatur}
\end{document}